\documentclass[a4paper,11pt]{article}
\usepackage{cite}
\usepackage{jheppub} 
\usepackage{etoolbox}
\patchcmd{\maketitle}{\@fpheader}{}{}{}
\usepackage[T1]{fontenc} 
\usepackage[utf8]{inputenc} 
\usepackage{microtype} 
\usepackage{lmodern} 
\usepackage{mdframed} 
\usepackage{tikz,tikz-3dplot}
\usetikzlibrary{decorations.pathmorphing,arrows.meta,calc}
\usepackage{mathtools}
\usepackage{verbatim}
\usepackage{graphicx}
\usepackage{cancel} 
\usepackage{todonotes} 
\usepackage[normalem]{ulem} 
\usepackage{soul} 
\usepackage{amsmath}
\usepackage{amsfonts}
\usepackage{amssymb}
\usepackage{stmaryrd}
\usepackage[mathscr]{euscript}
\usepackage{mathtools}
\usepackage{mathrsfs}
\usepackage{multicol}
\usepackage{mleftright} \mleftright
\usepackage{tensor} 
\usepackage{braket} 
\usepackage{mathrsfs} 
\usepackage{bbold}
\usepackage{color}
\usepackage{braket}
\usepackage{slashed}
\usepackage{hyperref}
\usepackage{verbatim}

\newcommand{\be}{\nopagebreak[3]\begin{equation}}
\newcommand{\ee}{\end{equation}}
\newcommand{\bee}{\nopagebreak[3]\begin{equation*}}
\newcommand{\eee}{\end{equation*}}
\newcommand{\ba}{\nopagebreak[3]\begin{eqnarray}}
\newcommand{\ea}{\end{eqnarray}}
\newcommand{\baa}{\nopagebreak[3]\begin{eqnarray*}}
\newcommand{\eaa}{\end{eqnarray*}}
\newcommand{\bal}{\nopagebreak[3]\begin{aligned}}
\newcommand{\eal}{\end{aligned}}

\newcommand{\bseq}{\nopagebreak[3]\begin{subequations}}
\newcommand{\eseq}{\end{subequations}\noindent}

\title{Thermodynamics of Swirling Black Holes and Kerr-Taub-NUT}

\author[1,2]{Ernesto Frodden}
\author[3]{Diego Hidalgo}
\author[1]{Francisco Manzano}
\affiliation[1]{Departamento de Física, Universidad Técnica Federico Santa María, Avenida España 1680, Valparaíso, Chile}
\affiliation[2]{CENTRA, Departamento de Física, Instituto Superior Técnico, Universidade de Lisboa, Av. Rovisco Pais 1, 1049, Lisboa, Portugal}
\affiliation[3]{Nordita, KTH Royal Institute of Technology and Stockholm University Hannes Alfvéns väg 12, SE-106 91 Stockholm, Sweden}

\emailAdd{efrodden@gmail.com}
\emailAdd{diego.hidalgo@.su.se}
\emailAdd{franciscoxmanzano@gmail.com}

\preprint{{\bf } }

\abstract{
We establish the first law of black hole thermodynamics for the recently discovered Kerr swirling (KS) and accelerated Kerr swirling (AKS) black hole solutions. We also revisit the first law for the Kerr-Taub-NUT (KTN) spacetime. Guided by Noether's original insight that global symmetries give rise to conservation laws, we employ and adapt the surface charge method for computing the associated charges (mass and angular momentum) in the presence of spacetime string-like defects (conical singularities) along the axes of both KS and KTN black holes. We obtain finite integrated charges in a quasi-local manner, without relying on the non-flat asymptotic regions characteristic of these spacetimes. We then establish a first law without introducing independent stringy terms unrelated to symmetries. For both KS and KTN, the quantities that play the role of the black hole entropy differ from the simple geometrical rule of one-fourth of the horizon area. This mismatch arises because we keep first-law charges associated with conservation laws. It indicates that standard formulas may not be naively used for black holes with string-like defects; they have a different topology and suggest the need for a new geometrical framework. 

In stark contrast, for the AKS spacetime, which can be regularized outside the singularity and lacks string-like defects, we find that the charges are consistent with a first law having the usual one-fourth of the area entropy. We first determine the mass using the Christodoulou-Ruffini formula and then check its agreement with the result obtained from the surface charge method.
}

\begin{document}
\maketitle
\flushbottom
\section{Introduction}

 \label{sec:intro}
Black holes offer a unique setting to explore the relationship between gravity and thermodynamics. Since the discovery of Hawking radiation, this connection has raised fundamental questions about the meaning of entropy, the nature of energy in gravitation, and the scope of black hole thermodynamics across different spacetime geometries. In this work, we address these questions by investigating black hole thermodynamics in spacetimes that lack an asymptotically flat region or contain string-like defects.

In recent years, there has been renewed interest in constructing new exact solutions of General Relativity, driven largely by solution-generating techniques such as the Ernst formalism and its Ehlers and Harrison transformations \cite{Ehlers:1957,Ernst:1967wx}. In this work, we study the thermodynamics of a recent family of black hole solutions: First, the Kerr black hole embedded in a swirling universe (KS)~\cite{Astorino:2022aam}. Recent studies of their properties appear in~\cite{Capobianco:2024jhe,Ouyang:2025qru,Capobianco:2025lxn,Barrientos:2026drd}. Second, its accelerated generalization (AKS), which adds an acceleration parameter $A$ \cite{Astorino:2022prj}. And third, the Kerr--Taub--NUT spacetime (KTN), which is not new but which we revisit in order to improve a previous work of some of the authors \cite{Frodden:2021ces}. All three contain Kerr black holes as a limit, which suggests a straightforward
extension of their thermodynamics. As we will see, this is not the case.

The standard description relies on the classic result of Hawking radiation that tells us that the temperature is related to the surface gravity, $T_{\rm H}=\kappa_{\rm H}/2\pi$, as well as the frequently cited area law for the entropy: $S=\text{Area}/4$.  Complementarily, the thermodynamic description usually relies on a presupposed asymptotically flat region, where idealized observers fix preferred time-like vectors at infinity and, with them, define a spacetime energy/mass. None of these spacetimes meet these conditions.

Moreover, KS and KTN display string-like defects along the axes~\cite{Astorino:2022aam,Misner:1963fr}: Conical singularities in KS and Misner strings in KTN. The AKS black hole, in contrast, can be freed of them \cite{Astorino:2022prj}. The strings introduce a technical problem, since naively computed charges depend on the integration surface, and a conceptual one, since it is unclear whether the standard formulas for temperature, angular velocity, and entropy remain correct.

This questioning has precedents. For spacetimes with a NUT parameter, the entropy was
found to differ from one quarter of the horizon area~\cite{Astefanesei:2004ji, Hawking:1998jf}. For KTN, imposing the standard formulas for $T$ and $\Omega$ over-determines the first law. In \cite{Frodden:2021ces} it was overcome by restricting the parameter space. As an alternative, and also keeping $S=\text{Area}/4$, the axes are regarded as additional Killing horizons, whose potentials and conjugate charges enter the first law as new terms~\cite{BallonBordo:2019vrn} (along the lines of \cite{Anabal_n_2019,Appels:2017xoe}).

Here, we attack these issues with the help of the surface charge method, as explained in \cite{Frodden:2019ylc}, based on the covariant phase space formalism \cite{Lee:1990nz,Iyer:1994ys,Wald:1999wa,Barnich:2001jy, Compere:2018aar}. The formalism is implemented quasilocally and does not require an asymptotic region;\footnote{An alternative method to deal with conical singularities from the action perspective can be found in \cite{Herdeiro:2009vd}.} it only requires the existence of Killing symmetries and the equations of Einstein satisfied in the region of interest. To handle the string defects, where Einstein's equations fail, we excise thin truncated cones around the defects from the integration domain. This restores the charges' independence from the surface. 

Then, the contributions of the cones are absorbed and required to define the mass and the angular momentum correctly; no additional conjugate pairs are introduced, unlike~\cite{BallonBordo:2019vrn}, where the tubes define independent conjugate charges. This choice defines our scope; we exclude explicit string contributions in the first law.

With a successful computation of the mass and the angular momentum for KS and KTN, we demand that the first law be satisfied and get $S$, $T$, and $\Omega$ from a set of differential equations plus the Kerr limit as a boundary condition. The resulting entropy does not match one quarter of the horizon area, and we stress that this follows from our choice to define charges via a conservation law. In this framework, the entropy is indeed a charge.

In KTN, on the one hand, the over-determination of~\cite{Frodden:2021ces} is now resolved without restricting the parameter space. On the other hand, the Misner-string terms of~\cite{BallonBordo:2019vrn} can be traced back to pieces of the charges computed in a particular surface choice (see \eqref{eq:decompo1} and \eqref{Ktermsandpotential}).

In AKS, the strings disappear when the parameters satisfy certain regularity relations~\cite{Astorino:2022prj}, and the acceleration is sustained by a gravitational spin--spin interaction. The computation of the mass gets involved, but the regularity motivates us to adopt $S=\text{Area}/4$ as a working hypothesis and then compute the AKS mass through the Christodoulou--Ruffini formula. We compare it with the surface-charge method in an $A$ expansion and find agreement. 

Again, for the AKS black hole, the $T$ and $\Omega$ differ from the standard ones, which can be understood by a normalization of time and a different choice of the rotating frame, whereas in KS and KTN the entropy itself departs from one quarter of the area, something that none of those choices can absorb.

The paper is organized as follows. Section~\ref{themethod} reviews the surface charge method and its adaptation to treat string-like defects. Sections~\ref{sec:KS}, \ref{sec:AKS}, and~\ref{section:revisitedKTN} deal with the KS, AKS, and KTN cases, respectively. We conclude in Section~\ref{sec:conclusions} with a summary of the results, their limitations, and open directions.

Notation warning. Symbols associated with each spacetime are defined within the corresponding section and should not be interpreted as referring to the same quantities elsewhere. We adopt this convention to avoid unnecessary subscripts. For example, $\xi$ denotes the Killing vector generating mass in each section, although its explicit expression may differ from one black hole to another.

%
\section{Surface charge method}\label{themethod}
In this section, we briefly review the surface charge method as prescribed in \cite{Frodden:2019ylc}. 
An exact symmetry, here equivalent to a Killing symmetry generated by a Killing vector field $\xi$, produces an on-shell conservation law equation: $d \boldsymbol{k}_\xi\approx 0$ (or equivalently, without form notation: $\partial_\mu k^{\mu\nu}_{\xi}\approx 0$). The integration of this equation over a spacetime volume $\Sigma$, with two disconnected boundaries $\partial \Sigma= \mathcal S \cup \mathcal S^{\prime}$, defines a {\it conserved} surface charge variation as
\begin{equation}
 \slashed{\delta} Q_{\xi}=   \oint_{\mathcal S} \boldsymbol{k}_\xi \,, 
\label{scintegral}
\end{equation}
the conservation can be equivalently stated as the property that for any other closed two-surface $\mathcal{S}^{\prime}$, obtained from a continuous deformation of $\mathcal S$, the previous expression produces the same $\slashed{\delta} Q_{\xi}$. For exact black hole solutions, this is true for surface deformations in any direction, in particular for radial directions. 
 In \eqref{scintegral} we use bold notation to indicate spacetime differential forms, namely the surface charge density is the two-form 
\begin{equation}
\label{surfacechargedensity}
\boldsymbol{k}_{\xi}  =\frac{1}{2} \Big( \delta \widetilde Q^{\mu \nu}_{\xi}-\widetilde Q^{\mu \nu}_{\delta\xi} + 2\xi^{[\mu} \Theta^{\nu]} \Big)\,\varepsilon_{\mu \nu \alpha \beta }dx^\alpha \wedge dx^\beta\,,
\end{equation}
the second term is required when the symmetry generators is allowed to depend on phase space variables.
For the Einstein-Hilbert theory, we have the following two contravariant tensor densities \cite{Frodden:2019ylc}
\begin{eqnarray}
  \widetilde Q_\xi^{\mu\nu} &=& -\kappa\sqrt{-g}\nabla^{[\mu}\xi^{\nu]}\,,\\
  \Theta^\mu(g, \delta_\xi g)  &=& \kappa \sqrt{-g}\left(\nabla_\alpha\nabla^{(\alpha} \xi^{\mu)}-\nabla^\mu\nabla^\alpha \xi_\alpha\right)\,. 
\end{eqnarray}
Here $\kappa = 1/(16\pi G)$.\footnote{We work in units where Newton's constant $G=1$. } Often $ \widetilde Q_\xi^{\mu\nu}$ is referred to as the Noether potential because the Noether current trivially satisfies $J^{\mu}_{\xi} = \partial_{\nu} \widetilde Q_{\xi}^{\mu\nu}$, and $  \Theta^\mu(g, \delta_\xi g) $ is the boundary term coming from an integration by parts in the derivation of Einstein's field equations.
See \cite{Frodden:2019ylc} for more details and the explicit formula derivation of $k^{\mu \nu}_{(\xi)} $ in terms of either metric, tetrad, or connection variables.

Here, we stress two aspects of \eqref{scintegral}. First, the slashed delta symbol in $ \slashed{\delta} Q_{\xi}$ means that we compute a one-form in phase space which is not necessarily exact. Remember that differential forms can also be defined over phase space. Exactness in phase space, namely $ \delta (\slashed \delta Q_\xi)=\delta^2Q_\xi=0$, is an extra condition required for the existence of a finite charge $Q_\xi$; this is sometimes called the {\it integrability condition}, and it is only fulfilled in particular cases or with very special assumptions \cite{Barnich:2007bf, Hajian:2015xlp}. Second, the integration is over any closed smooth surface $\mathcal S$ on which the conservation law holds. As we will see, string-like singularities violate the smoothness requirement, so we must correct the integration region accordingly. 

Indeed, for spacetime with string-like defects piercing the surface $\mathcal S$, a direct computation of \eqref{scintegral} may fail as it produces a quantity dependent on the surface $\mathcal S$ (in practice, it becomes dependent on the radius of the chosen sphere). This is because conservation breaks down at the defects: $d \boldsymbol{k}_\xi\neq 0$, and the fields are not smooth solutions of Einstein's equation there. 

To solve the problem, we exclude a small region containing the strings, compute the surface charges, and then take the limit as that region shrinks around the strings. In other words, we integrate the conservation law equation over a volume without the strings, where it holds. As a consequence, the result does not decouple into the usual two integrals, one over each closed surface $\mathcal S^\prime$ and $\mathcal S$. Instead, \eqref{scintegral} gains extra contributions.  

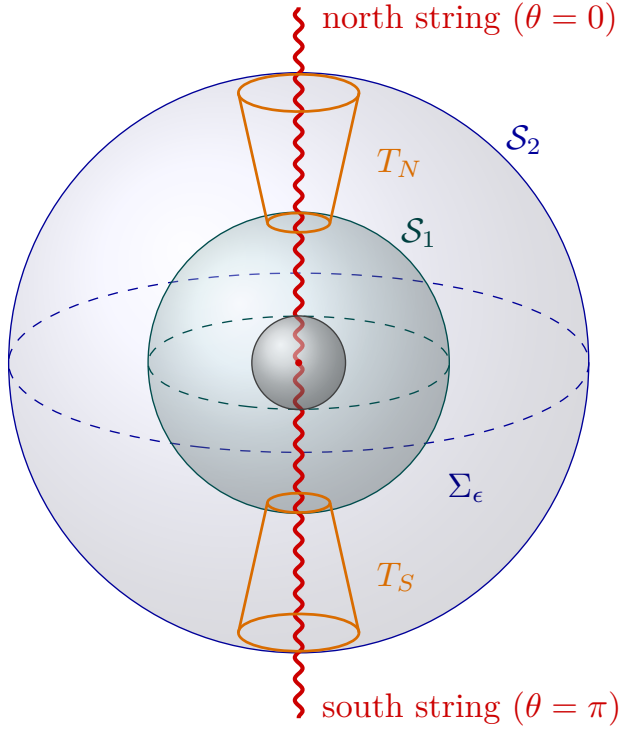
\begin{figure}[t]
\centering
\resizebox{0.7\textwidth}{!}{%
\tdplotsetmaincoords{72}{110}
\pgfmathsetmacro{\rIn}{1.35}   
\pgfmathsetmacro{\rOut}{2.6}    
\pgfmathsetmacro{\rBH}{0.42}    
\pgfmathsetmacro{\openang}{12}    
\pgfmathsetmacro{\zIn}{\rIn*cos(\openang)}  \pgfmathsetmacro{\pIn}{\rIn*sin(\openang)}
\pgfmathsetmacro{\zOut}{\rOut*cos(\openang)}  \pgfmathsetmacro{\pOut}{\rOut*sin(\openang)}
\begin{tikzpicture}[tdplot_main_coords,scale=1.15,>={Stealth[length=2mm]},
  lab/.style={font=\small},
  str/.style={very thick,red!80!black,decorate,decoration={snake,amplitude=0.4mm,segment length=2.2mm}}]

\begin{scope}[tdplot_screen_coords]
  \shade[ball color=blue!20,opacity=0.22] (0,0) circle (\rOut);
  \draw[blue!60!black] (0,0) circle (\rOut);
\end{scope}
\tdplotdrawarc[tdplot_main_coords,blue!60!black,dashed]{(0,0,0)}{\rOut}{0}{360}{}{}

\begin{scope}[tdplot_screen_coords]
  \shade[ball color=teal!30,opacity=0.35] (0,0) circle (\rIn);
  \draw[teal!60!black] (0,0) circle (\rIn);
\end{scope}
\tdplotdrawarc[tdplot_main_coords,teal!60!black,dashed]{(0,0,0)}{\rIn}{0}{360}{}{}

\draw[str] (0,0,0) -- (0,0,3.35);
\draw[str] (0,0,0) -- (0,0,-3.35);

\begin{scope}[tdplot_screen_coords]
  \shade[ball color=gray!55,opacity=0.55] (0,0) circle (\rBH);
  \draw[gray!50!black] (0,0) circle (\rBH);
  \fill[red!80!black] (0,0) circle (0.9pt);
\end{scope}

\foreach \s/\col in {1/orange, -1/orange}{
  \foreach \ph in {110,290}{
    \draw[\col!85!black,thick] ({\pIn*cos(\ph)},{\pIn*sin(\ph)},{\s*\zIn}) -- ({\pOut*cos(\ph)},{\pOut*sin(\ph)},{\s*\zOut});
  }
  \tdplotdrawarc[tdplot_main_coords,\col!85!black,thick]{(0,0,{\s*\zIn})}{\pIn}{0}{360}{}{}
  \tdplotdrawarc[tdplot_main_coords,\col!85!black,thick]{(0,0,{\s*\zOut})}{\pOut}{0}{360}{}{}
}

\node[lab,blue!60!black] at (0,2.15,2.34) {$\mathcal S_2$};
\node[lab,teal!50!black] at (0,1.15,1.35) {$\mathcal S_1$};
\node[lab,orange!85!black,anchor=west] at (0,0.62,1.95) {$T_N$};
\node[lab,orange!85!black,anchor=west] at (0,0.62,-1.95) {$T_S$};
\node[lab,red!80!black,anchor=west] at (0,0.1,3.25) {north string ($\theta=0$)};
\node[lab,red!80!black,anchor=west] at (0,0.1,-3.25) {south string ($\theta=\pi$)};
\node[lab,blue!50!black] at (0,1.6,-1.0) {$\Sigma_\epsilon$};
\end{tikzpicture}%
}
\caption{Integration region for the surface charge in the presence of string-like defects. The spacelike volume $\Sigma_\epsilon$ lies between the spheres $\mathcal S_1$ and $\mathcal S_2$ ($r_1<r_2$) and outside the
thin tubes $T_N$ and $T_S$ enclosing the north and south strings (red), taken as the truncated cones $\theta=\epsilon$
and $\theta=\pi-\epsilon$.}
\label{fig:tubes}
\end{figure}

To see how they emerge, let us analyze them in more detail. To fix ideas, consider a time- and axially symmetric spacetime where the string defects lie all along the axis. We can set an origin so we can refer to a string defect line in the north and one in the south of the spacelike volume; see Figure \ref{fig:tubes}. Consider the surface charge density $\boldsymbol{k}_{\xi}$ for a Killing field $\xi$. We integrate the three-form $d\boldsymbol{k}_{\xi}$ over a volume $\Sigma_\epsilon$ limited by four surfaces: Two spherical shells centered at the origin, say $\mathcal S_1$ and $\mathcal S_2$, with radii $r_1<r_2$, and two small tubes $T_S$ and $T_N$. For computational convenience, we actually consider truncated cones, excluding the strings from the volume, $\partial \Sigma_\epsilon=\mathcal S_1\cup \mathcal S_2\cup T_S\cup T_N$.\footnote{By $d\boldsymbol{k}_{\xi}\approx 0$, any other choice of tubes gives the same result in the $\epsilon\to 0$ limit.} The south truncated cone $T_S$ has a small opening angle, $\theta =\pi -\epsilon$, pointing towards the south, and the north truncated cone $T_N$ has a small opening angle, $\theta =0+\epsilon$, pointing towards the north. We integrate $d\boldsymbol{k}_{\xi}\approx 0$ over $\Sigma_\epsilon$
\be
\label{conservationlawTaub}
0= \int_{\Sigma_\epsilon}\!\! d \boldsymbol{k}_{\xi} =\int_{\partial \Sigma_\epsilon}\!\!  \boldsymbol{k}_{\xi}=\int_{\mathcal S_1}\!\! \boldsymbol{k}_{\xi} -\int_{\mathcal S_2}\!\! \boldsymbol{k}_{\xi}   + \int_{T_S}\!\!\! \boldsymbol{k}_{\xi}  +\int_{T_N}\!\!\! \boldsymbol{k}_{\xi}\,.
 \ee
In the absence of string-like singularities, the last two cone terms vanish in the shrinking limit $\epsilon\!\to\! 0$ and we recover the usual expression for the charge \eqref{scintegral}. However, for spacetimes such as the ones addressed in this paper, strings are present, and the cone contributions are finite in the limit $\epsilon\!\to\! 0$.

Although no general closed-form expression is available, the radial integration over each truncated cone can be formally done. Each cone term decouples into two terms associated with the circles found as the intersection of the cone with the spheres
\bseq\label{tubos}
\ba
\int_{T_S} \boldsymbol{k}_\xi & =& K_{\xi}\big|_{r=r_1; \,\theta=\pi-\epsilon} - K_{\xi}\big|_{r=r_2; \,\theta=\pi-\epsilon}\,,\\
\int_{T_N} \boldsymbol{k}_\xi & =&  K_{\xi}\big|_{r=r_1; \, \theta=\epsilon} - K_{\xi}\big|_{r=r_2; \,\theta=\epsilon}\,.
\ea
\eseq
Then, we take the limit $\epsilon\to 0$ and reorder \eqref{conservationlawTaub} with terms associated with each sphere to get an expression for the surface charge that is defined exclusively at a sphere $\mathcal S=\mathcal S_1$
\be\label{chargetaubnutgeneral}
\slashed{\delta}Q_\xi =\slashed{\delta}Q_\xi^{\text{(sphere)}}+\slashed{\delta}Q_\xi^{\text{(cones)}} = \oint_{\mathcal S} \boldsymbol{k}_{\xi} + K_{\xi}\big|_{\rm South} +   K_{\xi}\big|_{\rm North}\,,
\ee
where ``South'' and ``North'' refer to $\theta=\pi$ and $\theta=0$.  This quantity is truly conserved, {\it i.e.} its value does not depend on the specific surface $\mathcal S$. We emphasize that the linearity property of the surface charges still holds with this improvement 
\begin{eqnarray}\label{eq:linearity}
    \alpha_1\slashed{\delta}Q_{\xi_1}+\alpha_2\slashed{\delta}Q_{\xi_2} =\slashed{\delta}Q_{\alpha_1\xi_1+\alpha_2\xi_2}\,,
\end{eqnarray}
where $\alpha_{1,2}$ are phase space functions. In practice, for stationary and axially symmetric spacetimes, part of the cone terms exactly compensates the radial dependence of the sphere term, while an extra part can complete the value of the computed surface charge.

Although we formally use a specific parametrization for the geometrical objects, the result does not depend on it. The conservation $d\boldsymbol{k}_{\xi}\approx 0$ ensures that before shrinking the truncated cones around the strings, we can deform the spheres or the cones and still have the same resulting surface charge. In particular, we stress again that \eqref{chargetaubnutgeneral} does not depend on the specific radius at which we compute it. 

Let us recap: We have shown how certain constructed $K$-terms solve the problem of a naive computation with \eqref{scintegral} when string-like defects cross the surface $\mathcal S$. For spacetimes without strings, the $K$-terms automatically vanish.  We emphasize that obtaining \eqref{chargetaubnutgeneral} relies on correctly integrating the local conservation law. This requires identifying the tubes to exclude defects.

We now apply this improved method to the Kerr swirling spacetime announced above.

%
\section{Kerr black hole on a swirling universe}
\label{sec:KS}

In the search for new solutions to Einstein's equations in vacuum, $R_{\mu\nu} =0$, the authors in \cite{Astorino:2022aam} found a new spacetime family describing four-dimensional rotating black holes embedded in what has been called a swirling universe. This family generalizes the Kerr black hole family by adding a new parameter, the swirling parameter $\jmath$. We call it Kerr swirling spacetime (KS). The swirling universe as a background has recently attracted attention in the community. Several of its properties have been studied in different circumstances: The geodesic motion \cite{Capobianco:2024jhe}, the analysis of the accretion disks \cite{Ouyang:2025qru}, or the motion of light-like particles in this spacetime \cite{Capobianco:2025lxn}. 

A pure swirling universe, without embedded black holes, is a vacuum solution of Einstein's equations, axially symmetric, and equipped with an odd-$\mathbb{Z}_2$ symmetry that allows for opposite rotational dragging in both hemispheres. The astrophysical relevance of these spacetimes remains an open question. Two far-away identical mass distributions that are counter-rotating and therefore dragging the spacetime can produce an intermediate region that may be well-described by this swirling universe metric. Furthermore, we can imagine a third massive, but smaller and compact object sitting in the middle. If this object is also rotating, that full configuration would be a spacetime region well-described by a Kerr black hole in a swirling universe.

In the Boyer-Lindquist coordinates, the KS metric is described by\footnote{For the sake of compactness, we use slightly different definitions in the following functions, with respect to the ones used in \cite{Astorino:2022aam}.} 
\begin{equation}\label{eq:kerrswirlingmetric}
{ds}^2 = - F^{-1} \Bigg[\frac{\Delta}{\mathcal A} {dt}^2 -\biggl(\frac{{dr}^2}{\Delta} + {d\theta}^2\biggr)\Bigg]+   \sin^2(\theta ) F \mathcal A \left(d\phi - \frac{\omega}{\mathcal A} dt\right)^2 \,,
\end{equation}
where $F^{-1}(r,\theta)$ and $\omega(r,\theta)$ are functions that can be expanded to a finite power series of $\jmath$
\begin{align}
F^{-1}  & = \chi_{(0)} + \jmath \, \chi_{(1)} + \jmath^2 \, \chi_{(2)} \,, \qquad
\omega = \omega_{(0)} + \jmath\, \omega_{(1)} + \jmath^2 \, \omega_{(2)} \,,
\end{align}
with 
\begin{subequations}
\begin{align}
\chi_{(0)} & = r^2 + a^2 \cos^2(\theta) \,, \\
\chi_{(1)} & = 4 \,a \,m \, \Pi (r,\theta) \cos(\theta)  \,, \\
\chi_{(2)} & = \frac{4 \,a^2 \,m^2 \,\Pi(r,\theta)^2\, \cos^2(\theta) + \mathcal A(r,\theta)^2 \,\sin^4(\theta)}{r^2 + a^2 \cos^2(\theta)} \,,
\end{align}
\end{subequations}
and 
\begin{subequations}
\begin{align}
\omega_{(0)} & = -2 a m r+\omega_{00} \,\mathcal A(r,\theta)   \,,\label{eq:omega0eq} \\
\omega_{(1)} & = 4\cos\theta\,\Delta(r)\Big[r^3-a^2\big(m+(m-r)\cos^2\theta\big)\Big] \,, \\
\omega_{(2)} & = 2am\Big[-3r^5-6mr^4-2a^2r^3-12ma^2r^2-12m^2a^2r+a^4r+2ma^4 \nonumber\\
&\quad -6r\,(r^2+a^2)\,\Delta(r)\cos^2\theta+\big(r^3-3a^2r+2ma^2\big)\,\Delta(r)\cos^4\theta\Big]  \,,
\end{align}
\end{subequations}
finally 
\begin{subequations}
\begin{align}
\Delta(r) & = r^2 - 2mr + a^2 \,, \\
\mathcal A (r,\theta) & = (r^2 + a^2)^2 - \Delta(r)  \, a^2 \sin^2\theta\,,  \\
\Pi(r,\theta) & = r^2(\cos^2\theta - 3) - a^2 (1+\cos^2\theta)\,.
\end{align}
\end{subequations}
Here, the quantities $m$ and $a$ are inherited from the Kerr seed solution, while $\jmath$ is the so-called swirling parameter. For $\jmath =0$ the above metric describes the Kerr spacetime, and for $a =0$ and $\jmath \neq 0$ the metric describes a Schwarzschild black hole in a swirling universe. Setting the black hole quantities to zero, $m=a=0$, recovers the swirling universe. Notice that the compact object possesses two horizons read off from equation $\Delta(r)=0$, at the usual radii $r_{\pm} = m \pm \sqrt{m^2-a^2}$. Furthermore, these spacetimes \eqref{eq:kerrswirlingmetric} are time- and axially symmetric, with $\partial_t$ and $\partial_\phi$ the respective Killing vectors.  

The $\omega_{00}$ parameter in \eqref{eq:omega0eq} accounts for a shift in the $\phi$ coordinate and can be absorbed; thus, we set it to zero below, i.e., $\omega_{00}=0$. Therefore, this family of black holes is spanned by three parameters: $m$, $a$, and $\jmath$; related to the mass, the angular momentum, and the swirling property, respectively. 

\subsection{Charges}
It has already been checked in \cite{Astorino:2022prj} that the spacetime is asymptotically non-trivial and suffers from conical defects along the axis. This can be observed directly by considering the ratio between the perimeter and radius of a small circle around the rotational axis at both axes \cite{Astorino:2022aam, DiPinto:2024axv}. Therefore, to compute the surface charge correctly, one has to account for the defects as explained in the previous section. 

\subsubsection{KS angular momentum}
We start with the surface charge associated with the axial symmetry, $\partial_\phi$. The computation goes smoothly and directly produces an integrable quantity, $ \delta(\slashed\delta Q_{\partial_\phi})=0$, explicitly
\begin{equation}
\label{axialKS}
\slashed\delta Q_{\partial_\phi} = \frac{1}{\left(1-\eta^2\right)^2}\,  \left( \delta (a m) +16 \, a^2 \,\jmath \, m^2 \, \left(\delta (a\, \jmath \, m) + a \, m \, \delta \jmath \right) \right),
\end{equation}
we introduced $\eta= 4 m a \jmath $, a parameter that controls the swirling correction. Indeed,  $\slashed\delta Q_{\partial_\phi}$ is written such that it is easy to check that here the Kerr angular momentum is recovered for $\jmath=0$. 

Surprisingly, the $K$-terms coming from string-like singularities, \eqref{chargetaubnutgeneral}, do not contribute to the surface charge expression. In fact, each $K$-term is zero on its own. Because of its phase space integrability, we can compute the actual charge, which is indeed the swirling-corrected angular momentum, $J\equiv Q_{\partial_\phi}$, given by 
\begin{equation}
\label{angularKS}
 J \ =\ \frac{a \, m}{1-\eta^2} \,.
\end{equation}

\subsubsection{KS mass}

Now we proceed to compute the surface charge associated with the time symmetry generator $\partial_t$, which traditionally is associated with the spacetime energy/mass. 

A couple of subtleties in the computation are worth explaining in detail. The spherical integral, $\int_{\mathcal S} k_{\partial_t}$, is in general difficult to compute. To accomplish it here, we first split the integrand $k_{\partial_t}$ into its different variations: Terms proportional to $\delta m$, $\delta a$, and $\delta \jmath$. For each of them, we perform a series expansion in the $\jmath$ parameter and do the sphere integration, $(\theta,\phi)$, term by term. In practice, we compute only a few terms of each series and then extrapolate them to get the sum. 

As an example consider the split $\boldsymbol{k}_{\partial_t}=\boldsymbol{k}_{\partial_t}^{(\delta m)}+\boldsymbol{k}_{\partial_t}^{(\delta a)}+\boldsymbol{k}_{\partial_t}^{(\delta \jmath)}$. The spherical integration for the term with $\delta m$ is  
\begin{eqnarray}
\label{expansion}
\int_{\mathcal{S}} \boldsymbol{k}_{\partial_t}^{(\delta m)}&=&2\pi\int_0^{\pi} \left(f_0(r,\theta)+f_2(r,\theta)\jmath^2+f_4(r,\theta)\jmath^4+f_6(r,\theta)\jmath^6+\cdots \right) d\theta\, \delta m\,, \\  
&=&\left(1+\frac{7}{2}\eta^2+\frac{13}{2}\eta^4+\frac{19}{2}\eta^6+...\right)\,\delta m-r\Bigg(\eta^2 +\eta^4 +\eta ^6+\cdots\Bigg) \frac{\delta m}{m}\,, \\
&=&\left(1+\frac{1}{2}\sum_{n=1}^{\infty}(6\,n+1)\,\eta^{2n}\right)\delta m-r\Bigg(\sum_{n=1}^{\infty}\eta^{2n}\Bigg)\frac{\delta m}{m}\,, \\
&=&\frac{(\eta^2+1)(\eta^2+2)}{2(1-\eta^2)^2}\, \delta m-r\frac{\eta^2}{1-\eta^2}\frac{\delta m}{m}\,.
\end{eqnarray}
In the first line, the computation in $\phi$ is trivially $2\pi$, and we do the expansion in $\jmath$ where only even powers of $\jmath$ survive. In the second line, with the definition $\eta= 4 m a \jmath $,  we assume we can exchange the integral and the infinite expansion, and we integrate the first four terms. Note the dependence on $r$; we write it separately. In the third line, as we compute only a few terms (the computation time grows fast), we can extrapolate a general term for the integrated series. The series are computable in the fourth line, with the assumption $\eta<1$, and we get a compact result. 

We perform this algorithmic process for each variation. Collecting the partial results, the full result for the spherical contribution to the surface charge is
\begin{eqnarray}
 \nonumber      \slashed{\delta} Q_{\partial_t}^{\text{(sph)}} &=&\frac{1}{2(1-\eta^2)^2}\left((\eta^2+1)(\eta^2+2)\,\delta m+\eta^2 (7-\eta^2)\frac{m}{a}\delta a+2\eta^2 (\eta^2+2)\frac{m}{\jmath}\delta \jmath \right)\\
       &&-\, r\, \frac{\eta^2}{1-\eta^2}\Bigg(\frac{\delta m}{m}+\frac{\delta a}{a}+\frac{\delta \jmath}{\jmath}\Bigg).
\end{eqnarray}

We observe there is an $r$-dependence, isolated as the last term in the previous expression. As expected, it is exactly canceled by computing the $K$-terms contribution \eqref{chargetaubnutgeneral}
\begin{equation}
   \slashed\delta Q_{\partial_t}^{\text{(cones)}} =r\frac{\eta^2}{1-\eta^2} \left(\frac{\delta m}{m}+\frac{\delta a}{a}+\frac{\delta \jmath}{\jmath}\right)=- \frac{r}{2}\delta \bigg( \log (1-\eta^2)\bigg)\,,    
\end{equation}
this restores the surface charge conservation, which stands for an $r$-independence of the charges. To obtain this compact result, we performed a similar expansion procedure as before.

Then, considering the sum of these two quantities, and rearranging a bit, we obtain
\begin{equation}
\label{timeKS}
\slashed\delta Q_{\partial_t} =\slashed\delta Q_{\partial_t}^{\text{(sph)}}+\slashed\delta Q_{\partial_t}^{\text{(cones)}}=\delta m -\frac{3 m \eta ^2 }{2 \jmath(1- \eta ^2)}\, \delta \jmath + 
 \frac{m}{4}\delta\left(\frac{6}{1-\eta^2} - \log(1-\eta^2)\right) \,,
\end{equation}
which is a quantity conserved in the sense that it does not depend on the closed two-surface we choose to compute it on, as long as it encloses the singularity. 

This surface charge, however, is not integrable, $\delta (\slashed\delta Q_{\partial_t})\neq 0$, so we cannot expect to define a charge directly with it.

Next, we define an integrable charge using a different Killing vector constructed as a combination of the two Killing vectors. We rely on the linearity property \eqref{eq:linearity} of the space of the surface charges with respect to the symmetry generators.  Indeed, the most general symmetry generator can be expressed as $\xi=\alpha\, \left(\partial_t +\beta \, \partial_\phi\right)$ with $\alpha=\alpha (m,a,\jmath)$ and $\beta=\beta (m,a,\jmath)$ two arbitrary functions of the parameter space. Then, a general surface charge $\slashed\delta Q_\xi$ is
\be\label{eq:massswirling}
\slashed\delta Q_\xi = \slashed\delta Q_{\alpha (\partial_t +\beta \, \partial_\phi)} = \alpha \, \left(\slashed\delta Q_{\partial_t} + \beta\,\slashed\delta Q_{\partial_\phi}\right)\,.
\ee
Now we can just replace in the right-hand side of \eqref{eq:massswirling} the already computed expressions in \eqref{timeKS} and \eqref{axialKS}. As we demand $\slashed\delta Q_\xi$ to be an integrable charge related to the mass, we name it $M\equiv Q_\xi$. In particular, it should be a function of the three parameters $M=M(m,a,\jmath)$. Then, in order for this mass to exist, the equation \eqref{eq:massswirling} defines three differential equations that should be fulfilled, one for each variation $\delta m$, $\delta a$, and $\delta \jmath$.  The coupled system is simply given by
\ba
\frac{\partial M}{\partial m} &= &\frac{\alpha}{(1- \eta^{2})^2} \left( 1+\frac{3}{2} \eta^2 + \frac{1}{2} \eta^{4} + \beta\, a\, \left( 1+\eta^2 \right) \right)\\
\frac{\partial M}{\partial a} &=& \frac{\alpha\, m}{(1- \eta^2)^2}  \, \left( 8 a \jmath^2 m^2 \, ( 7- \eta^2)  +  \beta \, (1+\eta^2) \right) \\
\frac{\partial M}{\partial \jmath} &=& \frac{32 \alpha\, a^2 \, \jmath\, m^3}{(1- \eta^2)^2}\,  \left(1 +\frac{1}{2} \eta^{2} +\beta \, a\right) \,.
\ea 
 The system is analytically solvable for the three unknown functions $M(m,a,j)$, $\alpha(m,a,j)$, and $\beta(m,a,j)$. The solution is unique given a boundary condition; we chose the natural one corresponding to the mass of the Kerr solution, $M(m,a,\jmath = 0)=m$, thus
\be
\alpha = \frac{1}{1- \eta^2}\,, \hspace{0.5cm}\beta = -\frac{3\eta^2}{2a}\,,
\ee
and the mass function is given by
\be
\label{mass_KS}
M= \frac{m}{1-\eta^2}\,.
\ee
This is a swirling-corrected Kerr mass, analogous to \eqref{angularKS}. We also write down the explicit Killing vector $\xi$ that generates this mass
\be
\label{killingKS}
\xi = \frac{1}{1-\eta^2} \, \left( \partial_t - \frac{3\eta^2}{2a} \, \partial_\phi \right)\,.
\ee
Thus, to make this charge integrable, interpreted as the black hole's mass, we needed to correct the time-symmetry-generating vector by adding the axial-symmetry-generating vector and rescaling. Indeed, this happens also in other black holes, for instance Kerr-(a)dS \cite{Caldarelli:1999xj} or a black hole in a magnetic (Melvin) background \cite{Astorino:2016hls}, and it is expected here due to the asymptotic rotational peculiarity of the swirling universe. 
Notice that this Killing vector is not timelike everywhere outside the event horizon. This is not an issue of our construction, but a generic feature of swirling backgrounds: Since the frame dragging of the swirling universe grows linearly along the axes, any combination $\partial_t+\Omega_0\,\partial_\phi$ with constant $\Omega_0$, including $\partial_t$ itself, becomes spacelike far enough from the black hole.

Note that the constant shift $-\frac{3\eta^2}{2a}\,\partial_\phi$ selects a rotating frame, and the factor $(1-\eta^2)^{-1}$ fixes a normalization, which are not derived from asymptotic observers in this spacetime. In particular, we learn that a global timelike character cannot serve as a criterion for selecting the mass generator.

In the regions where it is timelike, the Killing vector in \eqref{killingKS} can be thought of as defining a {\it canonical time} through $\partial_{\tau}=\xi$: A $\tau$-time-translation defines a mass charge.  

\subsection{First law}
Next, we explore how far we can go with the obtained charges to reconstruct a first law of black hole mechanics for this system. But first, two preliminary caveats that take these swirling black holes out of the ordinary. First, we reiterate that the KS spacetime is a pathological solution, as it contains string-like singularities extending up to infinity. Second, the asymptotic structure of swirling universes has a rotational dragging, which is angular-dependent and divergent.\footnote{Explicitly, consider an observer with proper time $\tau$: $\partial_\tau=\sigma(\partial_t+\Omega\partial_\phi)$ with zero angular momentum (ZAMO) $L\equiv g_{\mu\nu}(\partial_\tau)^\mu(\partial_\phi)^\nu =0$. For a large $r$, we have 
\begin{equation}
\Omega(r,\theta)=- \left.\frac{g_{t\phi}}{g_{\phi\phi}}\right|_{r\gg 1} \approx r\frac{\jmath}{2}\, g_-(\theta)-m\jmath\, g_+(\theta)  +\mathcal{O}\left( r^{-1}\right)\,, 
\end{equation}
where $g_\pm(\theta)=\eta  \left(\cos^4\theta-6 \cos^2\theta\pm 3\right)+8 \cos\theta$. Therefore, it cannot be static, and its angular velocity diverges.} 
 Thus, treating these black holes as true thermodynamic systems remains challenging. In particular, we lack a simple asymptotic region where we can place static observers and imagine them throwing particles to interact with the black hole. That leaves us without one of the most compelling pictures for devising these objects as thermodynamic systems.        

Having said that, and with the experience of other black hole systems with similar issues \cite{Astorino:2016hls,Astorino:2023ifg,Frodden:2021ces}, we can directly ask the first law to be fulfilled relying on quasi-local treatment. For this phase space, it stands for finding the remaining pieces of the equation
\be\label{eq:firstlawimposed}
\delta M = T\, \delta S + \Omega \, \delta J\,,
\ee
where, again, $\delta$ means variations on the parameters space spanned by $m$, $a$, and $\jmath$; and $T$, $S$, and $\Omega$ have to be determined. Usually, one invokes further physical arguments, but surprisingly, here the system is fully determined once the Kerr limit is imposed as a boundary condition, because the parameter-space dimensions match those of the unknown quantities, $T(m,a,\jmath)$, $S(m,a,\jmath)$, and $\Omega(m,a,\jmath)$. 

Then, the solution is
\ba
\label{entropy_KS}
S(m,a,\jmath) &= & \pi \left( a^2+ \frac{1}{(1-\eta^2)^2}\left(m+\sqrt{m^2-(1-\eta^2)^2a^2} \right)^2 \right)\\
T(m,a,\jmath) &= & \frac{1-\eta^2}{4\pi a^2 m}\, \left( a^2 - \frac{m}{(1-\eta^2)^2} \, \left(m-\sqrt{m^2 - (1-\eta^2)^2 \, a^2} \right)\right)\\
\Omega(m,a,\jmath) &= & \frac{1}{2am} \left(m - \sqrt{m^2 -(1-\eta^2)^2\, a^2} \right)\,.
\ea
By construction, all these functions satisfy that for $\jmath=0$ they reduce to the standard thermodynamic quantities of the Kerr black hole. They are therefore a good candidate for generalizing Kerr black hole thermodynamics to KS black holes. Note that the obtained intensive thermodynamic quantities $T$ and $\Omega$ differ from the standard ones computed for regular black holes as $T_{\rm H}=\kappa_{\rm H}/2\pi$ and $\Omega_{\rm H}=\left.-g_{t\phi}/g_{\phi\phi}\right|_{r_+}$. With our orientation, $J=+Q_{\partial_\phi}$ while the horizon rotates towards negative $\phi$, so $\Omega$ should be compared with $-\Omega_{\rm H}$; indeed, for $\jmath=0$ one has $\Omega=-\Omega_{\rm H}=a/(r_+^2+a^2)$, whereas for $\jmath\neq0$ they differ.

If one prefers to retain the standard definitions of $T_{\rm H}$, $\Omega_{\rm H}$, and $S=A/4$, then additional terms would be required to satisfy the first law. However, these terms would not arise from conservation laws and therefore could not be identified as charges. In this case, the entropy would likewise not be a charge associated with a Killing vector. Our point is that these standard quantities require substantial revision for black hole spacetimes with defects, just as they are already known to be modified when the spacetime is not asymptotically flat \cite{Caldarelli:1999xj,Gibbons:2004ai,Appels:2016uha,Anabalon:2018qfv}.

Now, introducing $\gamma\equiv 1-\eta^2$, and $\tilde a\equiv\gamma a$ we can rewrite all previous quantities, plus the mass \eqref{mass_KS} and the angular momentum \eqref{angularKS}, in a more compact and suggestive way 
\ba
\label{explicit_escalation}
M\gamma &= & m\\
S\gamma^2 &= & \pi\left( \tilde a^2+\left(m+\sqrt{m^2-\tilde a^2} \right)^2 \right)\\
T/\gamma &= & \frac{1}{4\pi \tilde a^2 m}\, \left( \tilde a^2 - m\left(m-\sqrt{m^2 - \tilde a^2} \right)\right)\\
J\gamma^2 &= & \tilde a m\\
\Omega/\gamma &= & \frac{1}{2\tilde am} \left(m - \sqrt{m^2 -\tilde a^2} \right)\,.
\ea
On the right, we have the usual Kerr expressions where $a$ is replaced by $\tilde a$, and on the left there is a scaling in terms of $\gamma$.

From the intensive quantities in the previous expressions, by construction, a Christodoulou-Ruffini formula \cite{Christodoulou:1971pcn}, with the proposed entropy, is satisfied
\be
\label{CHR_KS}
M^2 = \frac{S}{4\pi} + \frac{\pi J^2}{S}\,.
\ee
This is the usual formula for a Kerr black hole, modified with $a\to \tilde a$ plus an independent $\gamma$-rescaling of the extensive quantities given by the left-hand sides of \eqref{explicit_escalation}, which indeed disappears from the formula. 

That is, for the Kerr swirling black hole we should consider a generalization of the Christodoulou-Ruffini formula in which the black hole area is replaced by the new entropy $S$ given by \eqref{entropy_KS}. This poses further physical questions. The fact that $S\neq \text{Area}/4$ is not a surprise, as it has already been noted in black hole systems with string-like defects, specifically in the Kerr Taub-NUT black hole. We revisit it in Section~\ref{section:revisitedKTN}. 

But remember that in the Kerr case, the area is directly related to the so-called reduced mass: The minimum mass a rotating black hole can reach through energy extraction via a Penrose process \cite{Christodoulou:1971pcn}. Does this mass-extraction process change due to the string-like singularities? We leave this physical interpretation for later studies. We keep in mind that singularities are spacetime pathologies, but they still might serve as proxies for physically meaningful spacetimes.  

Indeed, as it was already suggested in \cite{Astorino:2022prj}, the strings can be thought of as actual spacetime ropes or struts that hold the black hole in place. They balance a spin-spin interaction that tends to accelerate the black hole along the axis. In other words, in contrast with usual fluid dynamics, a purely gravitational effect is that a spinning object interacts with a rotating background even if the spin of the object is aligned with the spacetime rotation axis. 

Here we can think of the Kerr black hole as a spinning object embedded in a swirling universe; the rotating spacetime and the black hole are aligned along the symmetry axis. The expectation is a force pushing along the symmetry axis, but the spacetime described so far does not allow for such an acceleration. Then, the self-consistency of the system suggests to us, through the appearance of the string-like defect, that some {\it extra} phenomena are in place to hold the black hole at that position.  

Fortunately, these string pathologies were recently overcome in \cite{Astorino:2022prj} by generalizing the previous KS spacetime. In the next section, we study the charges and the first law of this more general solution. It accounts for an acceleration arising from the spin-spin interaction; this solution allows the previously studied black hole to accelerate.

\section{Accelerating Kerr black hole on a swirling universe}
\label{sec:AKS}
The accelerating Kerr black hole in a swirling universe (AKS) is an even more general vacuum solution of general relativity. It further generalizes the KS of the previous section by introducing a new parameter that controls the acceleration along the symmetry axis. To present the metric and some basic properties, we follow \cite{Astorino:2022prj}. In coordinates $(t,r,x=\cos\theta,\phi)$, the metric reads
\begin{equation}\label{eq:metricAKS}
ds^2=\frac{|\Lambda|^2\,\Sigma}{\Gamma^2}\left[-\frac{\Delta_r\,P}{\Xi}\,dt^2+\frac{dr^2}{\Delta_r}+\frac{dx^2}{(1-x^2)P}\right]
+\frac{(1-x^2)\,\Xi}{\Gamma^2\,\Sigma\,|\Lambda|^2}\,\big(\Delta_\phi\,d\phi-\omega\,dt\big)^2\;
\end{equation}
with the metric functions given by
\begin{equation}
\Sigma=r^2+a^2x^2,\qquad \Gamma=1-Arx,\qquad P=1-2Amx+a^2A^2x^2\,,
\end{equation}
\begin{equation}
\Delta_r=(1-A^2r^2)(r^2-2mr+a^2)\,, \qquad \Xi=(a^2+r^2)^2P-a^2(1-x^2)\,\Delta_r \,,
\end{equation}
the swirling dressing
\begin{equation}\label{eq:Lambda}
|\Lambda|^2=\big(1+\jmath\,h_0\big)^2+\jmath^2f_0^2,\qquad
f_0=-\frac{(1-x^2)\,\Xi}{\Gamma^2\,\Sigma}\,,
\end{equation}
\begin{equation}\label{eq:h0}
h_0=\frac{2am}{\Gamma\,\Sigma}\Big\{r^2\big[x^3-3x+Ar(1+x^2)\big]-a^2\big[x^3+x+A(r-3rx^2)\big]\Big\}\,,
\end{equation}
and the rotation function
\begin{equation}\label{eq:omega}
\omega=\omega_{01}+\frac{\omega_0+\jmath\,\omega_1+\jmath^2\,\omega_2}{\Xi}\,,
\end{equation}
where $\omega_{0,1,2}(r,x)$ are explicitly given in Appendix~\ref{app:omega}. The parameters $(m,a,A,\jmath, \Delta_\phi)$ are associated with the mass,
rotation, acceleration, swirling, and conicity, respectively. The parameter $\Delta_\phi$ is artificially introduced to ease the control of the conical defect strength, and it is equivalent to changing the range of the $\phi$ coordinate. The parameter
$\omega_{01}$ controls the rotation of the coordinate frame; it does not change the local geometry or the charges, but it will be conveniently fixed to simplify a check at the end of this section. Thus, in principle, this AKS black hole family has five free parameters: $m$, $a$, $A$, $\jmath$, and $\Delta_\phi$. However, we will see in a moment that regularity fixes two of them in terms of the others, reducing the free parameters to three. 

Note some limits. This metric has the new parameter $A$, interpreted as the black hole acceleration. A way to see this acceleration effect is to take the $m,a,\jmath\to 0$ limits so that the resulting metric reduces to a Rindler spacetime in spherical coordinates. For $\jmath=0$ one has $|\Lambda|=1$ and \eqref{eq:metricAKS} is the accelerating Kerr metric in its standard
form \cite{Plebanski:1976gy,Hong:2004dm}, while, for $\jmath=a=0$ the metric reduces to the usual accelerated Schwarzschild black hole, known also as the C-metric (see the review \cite{Griffiths:2006tk} and references therein). For $A=0$ ($\omega_{01}=0$ and $\Delta_\phi=1$) it is the KS black hole with $\omega_{00}=0$, and for $A=\jmath=0$ and $\Delta_\phi=1$ it is a Kerr black hole.\footnote{The choice $\omega_{01}=0$ is the frame adapted to the $A\to0$ limit. Here we stress that we have not yet imposed the regularity conditions of the AKS metric, so $(m,a,A,\jmath,\Delta_\phi)$ are independent parameters. In Section~\ref{sec:AKS_mass_check}, once regularity is imposed, we use the freedom in $\omega_{01}$ to simplify the mass check. This is the reason for the two different choices of $\omega_{01}$, which are related by a rotation of the frame and describe the same geometry.}

Similar to the accelerated Kerr black hole, this spacetime has three horizons defined by the zeros of $\Delta_r=0$, namely, two at $r_\pm=m\pm\sqrt{m^2-a^2}$ and a non-compact one at $r_A=1/A$ identified as the Rindler horizon. We are interested in the parameter domain where the Rindler horizon lies outside the black hole event horizon, $r_+<r_A$. 

As expected, the metric \eqref{eq:metricAKS} contains conical defects, string-like singularities along the south and north axes, which, in contrast with the KS case, can be cured thanks to the inclusion of the parameter $A$. The regularization requires that when considering a small loop around the symmetry axis, the ratio of its perimeter to its radius should be $2\pi$, as the loop shrinks \cite{Astorino:2022prj}. This imposes two independent conditions along the axes, one for the north and one for the south semi-axis. We can solve both conditions simultaneously, effectively reducing the dimension of the parameter space. Keeping the parameters $m$, $a$, and $A$; the solution for the others is
\begin{eqnarray}
\label{regularity_conditions}
    \jmath &=& \frac{1 + a^2A^2 \pm \sqrt{(1+a^2A^2)^2-4A^2m^2}}{8aAm^2}\, ,\\
   \label{Deltaphi}   \Delta_\phi &=& \frac{1 + a^2A^2 \pm \sqrt{(1+a^2A^2)^2-4A^2m^2}}{2A^2m^2} \,.
\end{eqnarray}
To choose the sign, we analyze the $A\to 0$ limit of the first equation. The plus sign implies a divergence $\jmath\to \infty$, on the contrary, the negative sign implies $\jmath\to 0$. We then choose the negative branch, as it yields a metric related to the Kerr metric.\footnote {We disregarded the positive branch because it is a less physically interesting limit, but it is still a regular black hole solution. Indeed, on the positive branch for $\Delta_\phi$ and $\jmath$ and then taking the $A\to0$ limit (which in this branch means to consider $\jmath\to\infty$) plus a proper coordinate rescaling, one finds a new vacuum solution, as explained in \cite{Astorino:2026kuv}, named the Kerr-Levi-Civita solution.} In this case, acceleration and swirling property can be simultaneously turned off in order to smoothly recover the Kerr black hole. With these choices, the solution is free of string-like singularities and, indeed, a generalization of the Kerr metric. It is interesting to note that for fixed $m$ and $a$, and a small acceleration, $A\ll 1/r_+$, the first expression reduces to 
\be
\label{spinspin}
mA\approx 4 m a \jmath\,,
\ee
which can be interpreted as a Newtonian force due to a spin-spin interaction term $4\jmath (ma)\sim 4 \jmath\, J $ \cite{Astorino:2022prj}. This is the limit of a small black hole compared with its acceleration, or equivalently, a far-away Rindler horizon compared with the size of the black hole. It is directly related to the pure gravitational effect that tells us that a spinning test body in a rotating spacetime feels a force.

Now, we proceed to compute the charges and unveil the corresponding first law of thermodynamics for this family of regular black holes spanned by the three parameters $m$, $a$, and $A$. 

\subsection{Charges and first law}
\subsubsection{AKS angular momentum}
We start with the angular momentum, the charge associated with the Killing vector $\partial_\phi$. Since the metric is regular on the axes, there are no string defects, and the last two terms of \eqref{chargetaubnutgeneral} vanish automatically in what follows. We only need to worry about the spherical contribution \eqref{scintegral}. 

The computation simplifies considerably if we look closely at equation \eqref{surfacechargedensity}; schematically, the surface charge density is $k \sim \delta \widetilde Q + \xi \Theta$. Consider the components $(\partial_\phi)^\mu=( 0,0,0,1)$; the integration over the arbitrary closed surface $\mathcal S$ can be performed on a sphere of constant $r$. A closer look at the coordinates and components shows that $\xi \Theta$ vanishes on $\mathcal S$. Since the integrand reduces to $k \sim \delta \widetilde Q$, the charge is trivially integrable in phase space, and the variation can be pulled out of the integral to get
\begin{equation}
J\equiv Q_{\partial_\phi} =  \oint_{\mathcal S} \widetilde Q^{\mu\nu}_{\partial_\phi}d\Sigma_{\mu\nu}\,,
\end{equation}
with $\widetilde Q^{\mu\nu}_{\partial_\phi}$ the Noether potential for the axial symmetry and $d\Sigma_{\mu\nu}=\frac{1}{2}\varepsilon_{\mu\nu\alpha\beta}dx^\alpha dx^\beta$ (with the orientation $\varepsilon_{tr\theta \phi}=+1$) the proper integration element for the $r$-constant sphere $\mathcal S$. We stress that the previous reasoning is quite general and can be followed each time we want to compute the angular momentum.\footnote{This is indeed the reason why the Komar integral 
\begin{equation*}
   \frac{1}{16\pi}\oint \sqrt{-g}\,\nabla^\mu \xi^\nu d\Sigma_{\mu\nu}=\frac{1}{32\pi} \oint \sqrt{-g} \nabla^\mu \xi^\nu \varepsilon_{\mu\nu\alpha\beta} dx^\alpha\wedge dx^\beta=-\frac{1}{2} \oint \widetilde Q^{\mu\nu}_{\xi} \varepsilon_{\mu\nu\alpha\beta} dx^\alpha\wedge dx^\beta\,,   
\end{equation*}
This gives the correct angular momentum, but it fails in general and must be replaced by a better method to compute, for instance, the mass.} 

We now describe the computation in two versions: first, a general approach exploiting conservation, and second, a special observation specific to the family solution that trivializes the computation.

We note that for the integration on the sphere, the one in $ \phi$ is trivial, but the one in $x=\cos\theta$ is not. To evaluate it, we use the fact that it should be $r$-independent (conservation), and therefore evaluate the integrand at $r\to 0$, and then integrate over $x$. At the limit, the expression simplifies considerably, and the integral becomes tractable. However, this is risky because it is unclear whether the limit commutes with integration. In fact, a naive integration over $x\in[-1,1]$ diverges due to $\sim 1/x$ terms. But because $1/x$ terms are odd and the integration interval is symmetric, they will cancel in a principal-value (symmetric limit) prescription, $\lim_{\epsilon\to 0}\int_{|x|>\epsilon}$, and then we simply subtract them under the assumption that a proper regularization is carried out for small $r$. The final result is a clean expression for the angular momentum  
\begin{equation}\label{eq:JaccKerrswirling}
J=  \frac{a \,m}{1-\eta^2}\Delta_\phi^2\,,  
\end{equation}
which reduces to KS result \eqref{angularKS} for $\Delta_\phi\to 1$. In fact, this angular-momentum calculation can be carried out either for general parameters and thus for a non-regular metric, or also with the regularized subset \eqref{regularity_conditions}, since in this case, for neither of them there is a contribution from the string defects. 

Notably, we can repeat the previous procedure by taking the limit $r\to\infty$ before integrating in $x$, and, reassuringly, the result is exactly the same. This suggests that exploiting $r$-independence, even as $r\to 0$ or $r\to \infty$, is a reliable trick.

Now we present an alternative, more direct approach based on a particular observation. For the metric \eqref{eq:metricAKS}, with the functions left implicit, the integral over the sphere reduces to
\be
\label{eq:J_integral}
J=\frac{1}{8\pi} \oint_{\mathcal S} \sqrt{-g} \nabla^{[r}(\partial_\phi)^{t]} dx d\phi=\frac{\Delta_\phi^2}{8}\int_{-1}^{1}\frac{f_0^2\,\Gamma^2}{|\Lambda|^4(1-x^2)P}\,\partial_r\omega\;dx\,.
\ee
The constant $\omega_{01}$ drops out immediately. Moreover, the integrand is a total $x$-derivative, namely
\be
\label{eq:total_derivative}
\frac{f_0^2\,\Gamma^2}{|\Lambda|^4(1-x^2)P}\,\partial_r\omega
=-\partial_x\!\left[\frac{h_0+\jmath\left(f_0^2+h_0^2\right)}{|\Lambda|^2}\right]\,,
\ee
which is the $x$-component of the relation $(2.3)$ of \cite{Astorino:2022prj}, applied to the Ernst potential (2.12) there.\footnote{Our swirling parameter is related to the one of~\cite{Astorino:2022prj} by $\jmath = -\jmath_{\text{Astorino}}$, cf.\ eq.~(4.4) here and eq.~(2.14) there; this is only a convention, and nothing else in the paper depends on it. In our variables, $\mathrm{Re}\,\mathcal{E} = f_0/|\Lambda|^2$, $\mathrm{Im}\,\mathcal{E} = [h_0 + \jmath(f_0^2 + h_0^2)]/|\Lambda|^2$ and $\rho = \frac{\sqrt{\Delta_r(1-x^2)P}}{\Gamma^2}$.} This also follows from the $t-\phi$ component of the vacuum Einstein equations.\footnote{The bracket is,
up to a sign, the potential of $\partial_\phi$. It is the imaginary part of the Ernst potential
$(f_0+ih_0)/\Lambda$, with $\Lambda=1+\jmath h_0-i\jmath f_0$, used in \cite{Astorino:2022prj} to
build the solution by an Ehlers transformation of the accelerating Kerr seed.} Then, the integral
only depends on the values of the bracket on the axes. There, $f_0=0$ because of the factor $(1-x^2)$,
while \eqref{eq:h0} gives $h_0(r,\pm \, 1)=\mp\, 4am$ for all $r$. Hence
\be
\label{eq:J_AKS}
J=\frac{\Delta_\phi^2}{8}\left[\frac{h_0}{1+\jmath h_0}\right]_{x=1}^{x=-1}
=\frac{am\,\Delta_\phi^2}{2}\left(\frac{1}{1-\eta}+\frac{1}{1+\eta}\right)
=\frac{am}{1-\eta^2}\,\Delta_\phi^2\,,
\ee
which smoothly generalises the Kerr swirling result \eqref{angularKS}, obtained for $A=0$. For $\jmath=0$, it reduces to the angular momentum of the accelerating Kerr black hole, $am\Delta_\phi^2$. The result \eqref{eq:JaccKerrswirling} is exact, to all orders in $A$ and $\jmath$, and manifestly independent of the radius of $\mathcal{S}$. Its derivation uses neither the regularity conditions
\eqref{regularity_conditions} nor the value of $\omega_{01}$. On the regular branch, $\eta=mA\Delta_\phi$. 

\subsubsection{AKS mass}

For the mass computation, we have no similar tricks. We could try to use the same strategy as in KS, \eqref{expansion}, an expansion in $\jmath=\jmath(m,a,A)$, but unfortunately our attempts to compute $\slashed\delta Q_{\partial_t}$ get stuck very quickly due to the complexity of the metric components. If we had it, we could then proceed in a similar way we did for the Kerr swirling case in \eqref{eq:massswirling}, to obtain the mass and even the Killing vector that generates it, from the equation $\delta M=\alpha (\slashed\delta Q_{\partial_t} +\beta \delta J)$, because again, here we have three parameters and three unknown functions. Furthermore, we could complete the remaining missing thermodynamic quantities, $S$, $T$, and $\Omega$, by solving them from the first law, again using the match between three unknowns and three parameters directly in the first law.

Unfortunately, this path is useless without an analytic computation of the surface charge associated with time symmetry: $\slashed\delta Q_{\partial_t}$. To find a way out, we choose a different path. Instead of insisting on a hard analytical computation, we phrase a conjecture and then compute the finite mass relying on it.

\vspace{0.3cm}

{\it {\bf Conjecture:} Regular black holes, without string defects, satisfy the standard Christodoulou-Ruffini formula.}

\vspace{0.3cm}

That is, we will take the formula \eqref{CHR_KS} for granted. That formula, originally proposed in \cite{Christodoulou:1971pcn} and easily checked for Kerr black holes, has also been checked in many other specific black hole examples. The formula has been generalized for 
an array of collinear accelerating black holes \cite{Astorino:2021rdg}, modified with the inclusion of a cosmological constant \cite{Caldarelli:1999xj}, or reinterpreted for non-regular scenarios as we did within the comments below \eqref{CHR_KS}.  

The relevance of the formula in the context of a black hole as a thermodynamic system at equilibrium is that it is indeed the so-called {\it fundamental relation of thermodynamics}, an equation that relates all extensive quantities present: $M(S,J)$. From it, the first law is a mere consequence of its varied version.

As far as the authors know, a general proof adapted to a generic black hole definition is not available; therefore, for present purposes we conjecture it. 

We refer to the {\it standard} Christodoulou-Ruffini formula above to say that the entropy in the formula is one fourth of the black hole horizon area. This should be the case precisely because we are dealing with a regular black hole: free of conical defects; the only curvature singularity is hidden behind the horizon. A direct calculation in the Kerr/CFT approach \cite{Astorino:2022prj} also supports this.

For the rest, we simply use the computed angular momentum from \eqref{eq:JaccKerrswirling} and the entropy as 
\begin{equation}
S= \frac{{\rm Area}}{4}=\frac{2\pi \, m\, r_+ \,\Delta_\phi}{1-A^2 r_+^2} \,,
\end{equation}
with regularity ensured by replacing both $\Delta_\phi$ and $\jmath$ with the negative branch of \eqref{regularity_conditions}, and then we obtain the square of the mass as given by
\begin{subequations}
\begin{eqnarray}
\label{eq:Chris-Ruffi}
M^2 &\equiv& \frac{S}{4\pi} + \frac{\pi \,J^2}{S}\\
 &=&
m^2\frac{2(1 - a^2 A^2)}{\left((1 + a^2 A^2)^2 - 4 A^2 m^2\right) \left(1 + a^2 A^2 + \sqrt{(1 + a^2 A^2)^2 - 4 A^2 m^2}\right)}\,,\\
&=& m^2 \frac{2 r_A^4 \left(r_A^2 - r_+ r_-\right)}{\left(r_A^2 - r_+^2\right) \left(r_A^2 - r_-^2\right) \left(r_A^2 + r_+ r_- + \sqrt{\left(r_A^2 - r_+^2\right)\left(r_A^2 - r_-^2\right)}\right)} \,.\\
&=& \frac{m^2\Delta_\phi^2}{\left(1 - \eta^2\right)^2}\left(1- \eta^2+2(\Delta_\phi - 1) \right)
\end{eqnarray} 
\end{subequations}

In the second line, we write it in terms of the mass, rotation, and acceleration AKS parameters:  $m$, $a$, and $A$. In the third line, we present it in terms of the three horizon radii present in the solution. For completeness, in the fourth line we present it more compactly in terms of the parameters $\eta=4 ma \jmath$ and $\Delta_\phi$ described before \eqref{regularity_conditions}.
As expected, the mass reduces to the Kerr mass $m$ for $A=0$; similarly $\eta=0$ and $\Delta_\phi=1$.\\

We now have a candidate for the mass by bypassing a difficult calculation. Furthermore, explicitly having the function $M(S,J)$, allows us to vary it to get the first law $\delta M=T\delta S+\Omega\delta J$, and read off the associated temperature $T$ and the angular velocity $\Omega$ as
\begin{subequations}
\begin{eqnarray}
T &\equiv&\frac{\partial M}{\partial S}= \frac{1}{2M}\left(\frac{1}{4\pi} - \frac{\pi J^2}{ S^2}\right)\,,\\
\Omega &\equiv&\frac{\partial M}{\partial J}= \frac{\pi J}{M\,S}\,,
\end{eqnarray}
\end{subequations}
by a simple replacement and further manipulations, we have for the temperature 
\begin{subequations}
\begin{eqnarray}
T &=&\alpha \, T_{\rm H}\\
\label{Tconradios}
&=&\frac{(r_A^2 + r_+ r_-)(r_+ - r_-)}{4\sqrt{2}\pi \, r_+ r_A^2 (r_+ + r_-)} \sqrt{ \frac{(r_A^2 - r_+^2) \left( r_A^2 + r_+ r_- + \sqrt{(r_A^2 - r_+^2)(r_A^2 - r_-^2)} \right) }{(r_A^2 - r_-^2)(r_A^2 - r_+ r_-)} }
\end{eqnarray}
where in the first line we express the result in terms of $T_{\rm H}$ and $\alpha$ given by 
\begin{eqnarray}
\label{TH}
    T_{\rm H}&=& \frac{\sqrt{-a^2 + m^2} \left(1 - A^2 \left(m + \sqrt{-a^2 + m^2}\right)^2\right)}{4\pi m \left(m + \sqrt{-a^2 + m^2}\right)}\,,\\
    \label{alpha}
    \alpha&=&(1 + a^2 A^2) \sqrt{ \frac{1 + a^2 A^2 + \sqrt{(1 + a^2 A^2)^2 - 4 A^2 m^2}}{2 (1 - a^2 A^2) \left((1 + a^2 A^2)^2 - 4 A^2 m^2\right)} }\,,
\end{eqnarray}
these are useful definitions for later purposes. In the second line of \eqref{Tconradios}, we expressed the parameters in terms of the three horizon radii, which makes it look slightly more compact. For the angular velocity, we have the expressions
\begin{eqnarray}
\Omega &=&\frac{a \left(1 + A^2 \left(a^2 - 2 m \left(m + \sqrt{m^2-a^2}\right)\right)\right) }{2\, m \left(m + \sqrt{m^2-a^2}\right) }\sqrt{\frac{1 + a^2 A^2 + \sqrt{(1 + a^2 A^2)^2 - 4 A^2 m^2}}{2(1 - a^2 A^2)}}\\
&=& \frac{ a \left(r_A^2 - r_+^2\right)}{2 r_A^2 r_+ (r_+ + r_-)} \left( \sqrt{\frac{(r_A + r_+)(r_A + r_-)}{r_A^2 - r_+ r_-}} + \sqrt{\frac{(r_A - r_+)(r_A - r_-)}{r_A^2 - r_+ r_-}} \right)\,,
\end{eqnarray}
\end{subequations}
 in terms of the parameters and the horizon radii, respectively. 
 
 We observe that the temperature and the angular velocity appearing in the first law differ from the standard formulas stemming from the metric, i.e. $\Omega_{\rm H}\equiv-\left.g_{t\phi}/g_{\phi\phi}\right|_H = \frac{\omega(r_+)}{\Delta_\phi}$ and $T_{\rm H}\equiv \kappa_{\rm H}/2\pi$, where surface gravity is computed with the null-at-the-horizon Killing vector $\partial_t+\Omega_{\rm H}\partial_\phi$. As in Section~\ref{sec:KS}, $\Omega$ is to be compared with $-\Omega_{\rm H}$. Moreover, $\Omega_{\rm H}=\omega(r_+)/\Delta_\phi$ shifts with the frame constant $\omega_{01}$, so this comparison is frame-dependent; the frame-independent statement is the mass generator \eqref{eq:massgenrator}. Note that the actual temperature is $\alpha$-proportional to $T_{\rm H}$ given by \eqref{TH}. This is not a surprise. The AKS spacetime has a non-flat asymptotic region; therefore, the time or the axial coordinate used to describe the metric does not have the usual asymptotic interpretation.
 
 

\subsubsection{Killing generator and consistency check for the mass}\label{sec:AKS_mass_check}

By directly calculating the mass using the Christodoulou-Ruffini formula, we missed identifying the actual Killing combination that generates it. In what follows, we show that we can indirectly construct it by reverse-engineering from the established first law. We also use this special Killing vector to perform a reassuring check with the surface charge method: We show an explicit match between the mass we obtain and the mass obtained from the surface charge method explained at the beginning, up to order $A^4$. 
                
The key ingredients are 1) linearity of the surface charge formula, 2) use of the null generator of the black hole horizon, and 3) specific equivalence between the surface charge and the variation of the area when computed over a bifurcation surface.

Area as a charge: The last ingredient is subtle but is the standard observation used in the classic work \cite{Wald:1993nt,Iyer:1994ys} to declare that {\it black hole entropy is a Noether charge}. In fact, given a null Killing vector that generates the horizon, there is always a particular rescaled one such that, over a bifurcation surface $\mathcal{B}$, its surface charge density is simply $k_{\chi}=\delta\widetilde Q_{ \chi}$. Furthermore, it is exactly related to the area as
\begin{equation}
\label{deltaarea}
    \delta \left(\frac{{\rm Area}}{4}\right) = \delta \int_{\mathcal{B}}\widetilde Q_{\chi} = \int_{\mathcal{B}} \delta\widetilde Q_{\chi} = \int_{\mathcal{B}} \boldsymbol{k}_{\chi}\,, 
\end{equation}
where the previous equation singles out a particular Killing vector
\begin{equation}
\chi=\frac{1}{T_{\rm H}}(\partial_t+\Omega_{\rm H}\partial_\phi)\,,  
\end{equation}
it is proportional to the Killing vector, which is also the horizon null generator. The proportionality constant is exactly the inverse of the temperature parameter defined from the surface gravity and expressed in \eqref{TH}. Note the geometrical nature of the previous statement: A strict connection between the area and exact symmetry generators at the bifurcation surface. The area is fixed; therefore, this equation fixes $1/T_{\rm H}$.

Consider now the first law, with the $T$ and $\Omega$ previously found, with the quantities we already know, but written as coming from the surface charges associated with their respective symmetry generators
\begin{subequations}
    \begin{eqnarray}
\label{killing_search}
\delta M &=& T \,\delta \left(\frac{{\rm Area}}{4}\right) + \Omega \, \delta J   \,,\\
   \int_{\mathcal{B}} \boldsymbol{k}_{\xi} &=& T \int_{\mathcal{B}} \boldsymbol{k}_{ \chi}+\Omega\, \int_{\mathcal{B}} \boldsymbol{k}_{\partial_\phi}\,, \\
  \int_{\mathcal{B}} \boldsymbol{k}_{\xi} &=& \int_{\mathcal{B}} \boldsymbol{k}_{\left(T\,\chi+\Omega\, \partial_\phi\right)}\,, \label{killing_search3}
\end{eqnarray}
\end{subequations}
In the second line, we write the extensive quantities $M$, Area, and $J$ as surface-charge integrals for the unknown Killing vector $\xi$ and the known Killing vectors $\chi$ and $\partial_\phi$. Conservation allows us to evaluate them on any closed two-surface, so we choose to effectively evaluate them on a bifurcation surface $\mathcal{B}$. This is relevant just for the area term, because in the same second line, we made implicit use of \eqref{deltaarea}.

In the third line \eqref{killing_search3}, we use the linearity of surface charges to identify the Killing field that generates the mass
\begin{eqnarray}\label{eq:massgenrator}
    \xi=T\chi+\Omega\partial_\phi=\alpha  \left[\partial_t+\left(\Omega_{\rm H}+\frac{\Omega}{\alpha} \right)\partial_\phi\right]=\alpha \left(\partial_t+\Omega_{\rm int}\partial_\phi\right)
\end{eqnarray}
where in the second equality we replace the definition of $\chi$ and $\alpha=T/T_{\rm H}$, see \eqref{alpha}. In the third equality we
introduce the internal angular velocity $\Omega_{\rm int}$. 

With the Killing vector generating the mass at hand, we go one step further and check that the mass we found matches the one obtained from the surface charge method, i.e., $\delta M = \slashed\delta Q_\xi$. To simplify the calculation, we now choose the remaining freedom in $\omega_{01}$; we remind that we can always choose the rotating frame by redefining the axial coordinate. This change in the metric also affects the surface charge $\slashed\delta Q_{\partial_t}$, but it leaves the overall $\slashed\delta Q_\xi$ unchanged.
Thus, for convenience set $\omega_{01}$ such that it exactly kills $\Omega_{\rm int}=0$. This simplifies the check, and with this new metric we should show that 
\begin{equation}    
\delta M=\slashed\delta Q_\xi=\slashed\delta Q_{\alpha \partial_t}=\alpha \,\slashed\delta Q_{\partial_t}\,.
\end{equation}
The LHS is known; in the RHS, $\alpha$ is known, but $\slashed\delta Q_{\partial_t}$ is the stumbling block at the beginning. We divide both sides by $\alpha$, consider the variations on both sides along the parameters to find terms proportional to $\delta m$, $\delta a$, and $\delta A$, and perform a series in $A$ to compare both sides. We find that all terms coincide up to order $A^4$; beyond that, the computation stalls.

This provides a nontrivial check that the mass obtained from the conjectures matches what one would obtain from the surface charge method if the computations were smoother. 
As a final remark, we note that the selection of $\omega_{01}$, such that $\Omega_{\rm int}=0$, with an $\alpha$-rescaling of the time coordinate, $\tau\equiv t/\alpha$, stands for an adaptation of our metric coordinates such that the mass is simply generated by a {\it canonical time} with associated Killing vector $\partial_\tau$, and a surface charge $\delta M=\slashed\delta Q_{\partial_\tau}=\oint_{\mathcal{S}} k_{\tau}$.

\section{Kerr Taub-NUT spacetime revisited: The origin of Misner-string terms}\label{section:revisitedKTN}

We briefly note that the strategy used for the Kerr swirling spacetime, without acceleration, equally works in the case of Kerr-Taub-NUT spacetime. This spacetime has a string-like defect, the Misner string \cite{Misner:1963fr}, and a three-dimensional parameter space: $m$, $a$, and $n$. For more details on the metric, see \cite{Frodden:2021ces} and references therein. 

On one hand, using the cones carefully to take into account the strings, we can directly compute and solve the integrated charges that are the mass and the angular momentum: $M=m$ and $J=(a-3Cn)m$ (details in \cite{Frodden:2021ces}). On the other hand, we can equally impose the first law equation \eqref{eq:firstlawimposed}, and solve it for $S(m,a,n)$, $T(m,a,n)$, and $\Omega(m,a,n)$, in a similar strategy to that in \eqref{entropy_KS}. Here, for Kerr-Taub-NUT spacetime, the result is
\ba
\label{KTN}
S & =& 2m\pi \, \left( m + \sqrt{m^2 - (a-3Cn)^2}\right)\label{eq:entropyKTN}\\
T&=& \frac{\sqrt{m^2-(a-3Cn)^2}}{4m\pi \, \left(m+ \sqrt{m^2 - (a-3Cn)^2} \right)}\\
\Omega &=& \, \frac{a-3Cn}{2m \, \left(m+\sqrt{m^2 - (a-3Cn)^2} \right)}\,.	
\ea
Let us remember that the parameter $C=\{-1,0,+1\}$ reflects three equally valid choices for the position of the strings, while $n$ is the NUT parameter. While $C$ can be locally absorbed by a coordinate transformation, and therefore assumed as a gauge choice, the global aspect of the $C$ parameter is not trivial; this becomes clear at the charge computation.

The results for the thermodynamic quantities obtained here are in contrast with the ones obtained in \cite{Frodden:2021ces}, as there the logic was to impose the {\it standard formula} for the temperature from the surface gravity $T_{\rm H}=\kappa_{\rm H}/2\pi$ and the naive angular velocity $\Omega_{\rm H}=-\left.g_{t \phi }/g_{\phi\phi}\right|_{r_+}$. This logic implied an overdetermination of the first law equation, which was resolved only by restricting the parameter space to $m=\alpha_0 \, n $ to obtain an integrable entropy (equation (14) in \cite{Frodden:2021ces}). In this new treatment, with equations \eqref{KTN}, the final black hole family still has three parameters, but the resulting temperature does not coincide with $T_{\rm H}$. From the equations, we can check that they are a simple deformation of the Kerr quantities $a\to (a-3Cn)$, which in fact is innocuous for the sub-case $C=0$. Then, in the same spirit as before, by construction, the Christodoulou-Ruffini formula holds. Again, the entropy $S\neq \text{Area}/4$, but as mentioned before, it might be expected for spacetimes with string-like defects \cite{Astefanesei:2004ji,Garfinkle:2000ms,Hennigar:2019ive}. Similarly, for $T$, the defects might play a global role. We leave this for future investigations \cite{Kerner:2006vu}.

\subsection{ A tale of the origin of the Misner string terms in the first law }
In our approach, for any radius $r$, the mass and angular momentum decompose as
\begin{equation}
\label{rdescomposition}
    \delta M = \mathcal{S}_t (r)+ K_t(r)\,, \hspace{1cm} \delta J = \mathcal{S}_\phi (r) + K_\phi (r)\,,
\end{equation}
where $\mathcal{S}_\xi (r) \equiv\oint_{\mathcal S_r}\boldsymbol{k}_\xi $ is the sphere integration and in $K_\xi(r)\equiv K^{{\rm north}}_\xi(r)+K^{{\rm south}}_\xi(r)$ we packed the sum of the south and north circle contributions as defined in \eqref{chargetaubnutgeneral}. We consider $\xi$ as either $\partial_t$ or $\partial_\phi$. Each term depends on $r$ but their sum is $r$-independent, yielding: $M=m$ and $J=(a-3Cn)m$.

In the asymptotic region, for $r\to \infty$, the $K$-terms of $\partial_t$ vanish; then the mass at infinity is completely given by the sphere term as $\delta M=\delta m$. However, for $\partial_\phi$, the sphere contribution to $\delta J$ along $\delta n$ grows linearly in $r$, and only with the $K$-terms does one get a finite charge. So, even at infinity, the $K$-terms are needed to compute the correct angular momentum.\footnote{For $C=0$, the sphere alone gives a finite angular momentum $J=am$. The linear $r$--divergence appears only for $C=\pm 1$. } \\ 
In the horizon region, $r_+=m+\sqrt{m^2+n^2-a^2}$, we consider the horizon generator $\tilde \chi = \partial_t + \Omega_{\rm H}\,\xi_\phi = \partial_t - \Omega_{\rm H} \partial_\phi$ with $\Omega_{\rm H} = g_{t\phi}/g_{\phi \phi}\Big|_{r_+}$.\footnote{In this section we follow the orientation of \cite{Frodden:2021ces}, in which the angular momentum is the charge of $\xi_\phi\equiv-\partial_\phi$, $J=Q_{\xi_\phi}$. Accordingly, $\Omega_{\rm H}$ has the opposite sign to the horizon angular velocity of Secs.~\ref{sec:KS}--\ref{sec:AKS}.} Using that $\tilde\chi$ is also a Killing vector, there is a surface charge associated, $\slashed\delta Q_{\tilde\chi}$, and because of the linearity of the surface charges we can write this as $\slashed\delta Q_{\tilde\chi}=\delta M+ \Omega_{\rm H} \delta J$. But now the subtle observation: $\slashed\delta Q_{\tilde\chi}$ has a sphere and $K$-terms decomposition as in \eqref{rdescomposition}, but as $\tilde\chi$ is also a horizon generator. Evaluated on the horizon, the sphere term is proportional to the variation of the area, that is 
\begin{equation}\label{eq:decompo1}
   \delta M+ \Omega_{\rm H} \delta J = T_{\rm H} \delta \Bigg(\frac{{\rm Area}}{4} \Bigg) + K_{\tilde\chi}(r_+) \,, \hspace{1cm} K_{\tilde\chi}(r_+) = K_t(r_+) +\Omega_{\rm H} \, K_\phi (r_+)\,.   
\end{equation}
where $T_{\rm H} = \kappa_H/2\pi$ is the Hawking's temperature obtained from the surface gravity $\kappa_H$ at the event horizon and ${\rm Area} = 4\pi(r_+^2+a^2+n^2-2Cn a)$. The remaining term $K_{\tilde\chi}(r_+)$ is not integrable in the phase space.

Equation \eqref{eq:decompo1} and the previous analysis help us to connect two different approaches, besides the one used in the current work: 
\begin{enumerate}
\item {\it Restricting the phase space}: In \cite{Frodden:2021ces}, the charges are computed with the surface charge method, and no extra terms associated with Misner strings are added to the first law. We take the standard intensive quantities as the main external inputs: the temperature $T_{\rm H}$ and the angular velocity $\Omega_{\rm H}$. Then we are obliged to deal with $K_{\tilde\chi}(r_+)$ in \eqref{eq:decompo1}. It can be shown that the entire RHS of \eqref{eq:decompo1} can not be written simply as a $T_{\rm H}\delta \tilde S$ for any $\tilde S$. The proposed way out in \cite{Frodden:2021ces} is to reduce the parameter space by setting $m=\alpha_0\, n$, where $\alpha_0$ is an arbitrary constant. Then, we can define an entropy that is indeed integrable in phase space, but it is not ${\rm Area}/4$.

\item {\it New conjugate pairs:} In \cite{BallonBordo:2019vrn,BallonBordo:2019tyh}, the axes where the Misner strings are located are regarded as Killing horizons, so that the remaining term in \eqref{eq:decompo1} is written in terms of the quantities
\begin{equation}
  \psi_{\pm} = \frac{1}{8\pi n(1\pm C )}\,, \quad N_{\pm} = -\frac{2\pi n^2(C \pm 1)^2(n\mp a)}{r_+} \,, \quad J_{B}= \frac{{\rm Area}}{4\pi}\frac{(a-Cn)}{2r_+}\,,
\end{equation}
where the $\psi_{\pm} $ are proportional to the surface gravities of the  Killing generators of the Misner horizons, and $N_{\pm}$ are called conjugate Misner gravitational charges. Note that these are not surface charges coming from an isometry of the spacetime. Instead, $J_B$ and all the quantities come from a reordering of the Komar integrals evaluated both at the horizon and at infinity. Within this approach, the first law takes the form
\begin{equation}
    \delta M = T_{\rm H} \,\delta S -\Omega_{\rm H}\,\delta J_B +\psi_+ \,\delta N_+ +\psi_{-}\, \delta N_{-}\,,
\end{equation}
where $M=m$ and $S={\rm Area}/4$, and $J_B=J+ J_s=m(a-3C n) + J_s$, i.e. the actual angular momentum coming from a conserved law plus 
\begin{equation}
    J_s = \frac{an^2(1+C^2)+C n (r_+^2-2n^2)}{r_+}\,.
\end{equation}
One can check that the sum of the $K$-terms at the horizon is exactly
\begin{equation}
\label{Ktermsandpotential}
  K_{\tilde\chi}(r_+)  = K_t(r_+)+ \Omega_{\rm H} K_{\phi}(r_+) = -\Omega_{\rm H} \delta J_s + \psi_+ \delta N_+ + \psi_- \delta N_-\,.
\end{equation}
Notice that for $C=\pm 1$ one of the $\psi_{\pm}$ diverges, but the product $\psi_\pm \delta N_\pm$ is finite. This corroborates that all the Misner string terms are exactly the $ K$-term contribution from the surrounding circles evaluated at the horizon.

\item {\it Surface charges and abandonment of $S={\rm Area}/4$}: In our current approach, following the spirit of Noether charges, both $M$ and $J$ are conserved charges, here $r$-independent, because of the inclusion of the $K$-terms. Then, the first law takes its standard form $\delta M = T \, \delta S + \Omega \,\delta J$, and no additional terms or restriction of the phase space are needed. But  $T$, $\Omega$, and the entropy do not coincide with $T_{\rm H}$, $\Omega_{\rm H}$, and ${\rm Area}/4$, respectively.
\end{enumerate}
We recap this tale as follows:  The origin of the Misner terms in approach 2 becomes clear from the way the quantities are defined in \cite{BallonBordo:2019vrn}: A single Komar identity for a special Killing vector, integrated between the horizon and infinity, is split into pieces assigned to different surfaces. The mass is evaluated at infinity, where the $K$-terms of $\partial_t$ vanish; the entropy and the temperature at the horizon; $N_\pm$ through the whole Misner tubes; and $J_B$ is the Komar integral of $\partial_\phi$ on the horizon. In the presence of Misner strings, the latter varies with the sphere where it is computed, so $J_B$ is an $r$--dependent charge. However, the angular momentum, the conserved charge of $\partial_\phi$, should be instead $r$-independent, and a surface charge calculation gives $J =(a -3C n)m$, which coincides with the asymptotic charge quoted in \cite{BallonBordo:2019vrn}.
As is clear from \eqref{eq:decompo1}, the mismatch with the standard thermodynamic quantities is packed in the $K_{\tilde \chi}(r_+)$, which in turn can be repacked and reinterpreted in the pairs $(\psi_\pm , N_\pm)$, plus $J_s$. 

To decide which thermodynamic description is correct, we should complement the previous results with a first-principles derivation either of the temperature or a deeper explanation for the entropy. We discuss these points further in the conclusion. Still, already with the tools we have at hand, respecting the conservation laws associated with the exact symmetries is a straightforward guiding principle, as we did in the present work.

\section{Conclusion}
\label{sec:conclusions}
In this paper, we used the surface charge method to compute the conserved charges of the following axisymmetric black hole spacetimes: the Kerr black hole in a swirling universe (KS), the accelerated Kerr black hole in a swirling universe (AKS), and the Kerr Taub-NUT black hole (KTN). Since none of these spacetimes possesses an asymptotically flat region and due to the presence of string-like singularities, the computation of the mass and angular momentum deserves careful treatment. In this sense, the surface charge method provides a powerful framework for a quasilocal treatment based only on a Killing vector $\xi$ and the conservation law $d\boldsymbol{k}_\xi \approx 0$.

We summarized the key steps for handling all these spacetimes. 
First, we only require Einstein's equations to hold for the surface charge conservation law. This is the on-shell condition. In particular, no boundary terms, e.g., the boundary Gibbons-Hawking-York term, are needed in this approach. 
Second, the string-like singularities mean that, in a naive approach, the charge expression depends on the choice of the closed surface, violating conservation. We cured this issue by excluding these spacetime defects by means of incorporating $K$-terms associated with the truncated cones (see Figure \ref{fig:tubes}). Then, the surface charge is truly conserved even though the integration surface is punctured. This careful choice of integration region was vital for the KS and KTN examples where the strings give non-trivial contributions. In contrast, in the AKS spacetime, a precise relationship between the phase-space parameters allows the defects to be removed simultaneously. Then, computing the charges reduces to integrating over a smooth closed surface. 

It is worth mentioning that in the absence of an asymptotically flat region for these spacetimes, in principle, there is no preferred normalization of the time translation nor a preferred non-rotating frame. Here we have shown that by demanding the integrability of the quantity $\slashed{\delta} Q_\xi$, a canonical time can be selected. This procedure works in all the studied cases. This result corroborates that the timelike nature of a Killing vector cannot always be used to compute the mass.\footnote {Although in the AKS case it was checked indirectly by first computing the mass through the Christodoulou–Ruffini formula.} Instead, integrability in phase space is a more relevant condition. 

The AKS spacetime is the most interesting example because it is regular. Here, a suitable relationship between the phase parameters removes both conical singularities simultaneously. The acceleration arises from the interaction between the black hole's spin and the swirling background's rotation. Due to this removal of the singularities, the $K$-terms vanish, and the first law takes the standard form with the entropy being one fourth of the event horizon area. The angular momentum acquires a closed form which coincides (up to orientation) with the Komar integral. We compute the mass from the Christodoulou-Ruffini formula and check it by identifying the right generating Killing vector and computing the corresponding surface charge.

In the revised Kerr-Taub-NUT example, the Misner strings give non-vanishing $K$-terms, and we need them to obtain charges without $r$-dependence. We work on the full phase space parameter $(m,a,n)$ and find that the mass and angular momentum are integrable, and the first law takes the standard form without introducing extra thermodynamic variables associated with the strings or reducing the phase space parameters. In particular, the entropy is fixed uniquely by imposing the first law and the continuity limit with the Kerr black hole, and it does not coincide with ${\rm Area}/4$. 

Several natural directions for future work remain. In the KTN case, we did not compute the temperature and entropy directly, but inferred them by demanding the first law together with the Kerr limit. An independent derivation, not relying on the first law itself, is therefore needed to confirm them. For the temperature, this amounts to studying a quantum field on the KTN background and reproducing the Hawking computation, e.g.\ through Bogoliubov coefficients. 
We expect the modes used to construct the quantum state to be sensitive to the global structure introduced by the Misner strings, so that the resulting temperature $T$ need not coincide with $T_{\rm H}$. The entropy is harder, since no first-principles derivation exists. Here we expect a geometric object such as the holonomy group to play a central role, as it distinguishes point-like singularities, which give rise to an area term associated with closed surfaces, from string-like ones, which give rise to new contributions associated with punctured surfaces. We leave both ideas for future work.

For the AKS black hole, as a new rich arena, several directions arise. For instance, it would be interesting to further analyze a spinning test particle in the pure swirling background, its spin-spin force, and compare it with the small acceleration limit for the black hole \eqref{spinspin}. Or, to check that the positive branch choice in the regularity condition \eqref{regularity_conditions} does not change the observable quantities such that they also describe the thermodynamics of the Kerr-Levi-Civita solution, which is obtained through a parameter limit on the branch we left aside. A natural application of these surface-charge techniques to black hole thermodynamics is the KS black hole extension, which includes electric and magnetic charges \cite{DiPinto:2025yaa}.

Finally, our determination of the AKS mass provisionally makes use of the conjecture that regular black holes satisfy the standard Christodoulou--Ruffini formula. We alternatively checked the result through the surface charge approach (up to a certain order), which further supports the conjecture. A general proof of the conjecture is missing; we leave this for future investigations.

\begin{acknowledgments}
The authors thank Davide Astesiano for his early contributions to this work, and Marco Astorino for useful discussions. DH also thanks the Institute for Theoretical Physics of Utrecht University for its hospitality, where part of the project was carried out. The work of D.H. was supported in part by the Icelandic Research Fund 2511228-051 and the European Union’s Horizon Europe research and innovation program under Marie Sk\l odowska-Curie Grant Agreement No. 101210016. Nordita is supported in part by NordForsk. 

AI assistance was used for a general grammar review, symbolic cross-checks, and improving a figure already presented in \cite{Frodden:2021ces}. The observation to alternatively compute $J$ in \eqref{eq:J_integral} was also a product of AI interaction. 

We provide a Mathematica notebook with calculations in the arXiv folder.

\end{acknowledgments}

\appendix
\section{The functions $\omega_0$, $\omega_1$, $\omega_2$ for accelerated Kerr black hole in swirling universe}\label{app:omega}
In equation~\eqref{eq:omega}, with $\Gamma=1-Arx$ and $\Delta_r=(1-A^2r^2)(r^2-2mr+a^2)$, the components of $\omega$ are written in powers of the
acceleration, so that the Kerr swirling functions \eqref{eq:kerrswirlingmetric} are recovered by setting $A=0$
\begin{align*}
\omega_0&=-2amr+2Aamx(a^2+r^2)+A^2\big[2amr^3-a(a^2+r^2)(r^2+a^2x^2)\big],\\[1mm]
\omega_1&=-\frac{2(1+a^2A^2)\,\Delta_r}{(1-A^2r^2)\,\Gamma^2}\Big[-2x\big(r^3-a^2(m+(m-r)x^2)\big)\\
&\qquad+A\big(r^4(1+x^2)+4mr^3x^2-a^2r^2(1-x^4)-4ma^2rx^2-a^4x^2(1+x^2)\big)\\
&\qquad+2A^2rx\big(a^2r^2+a^4x^2-mr^3(1+x^2)\big)\Big],\\[1mm]
\omega_2&=\frac{2am}{\Gamma^3}\sum_{n=0}^{5}A^n\,\mathcal V_n\,,\\[1mm]
\mathcal V_0&=-3r^5-6mr^4-2a^2r^3-12ma^2r^2-12m^2a^2r+a^4r+2ma^4\\
&\qquad-\big[6r(r^2+a^2)x^2-\big(r^3-3a^2r+2ma^2\big)x^4\big](r^2-2mr+a^2),\\[1mm]
\mathcal V_{1}&=4r^{6}x(3+x^{2})+2mr^{5}x(9+2x^{2}-x^{4})+4m^{2}r^{4}x(3-6x^{2}+x^{4}) \\
&\qquad +3a^{2}r^{4}x(1+6x^{2}+x^{4})+4a^{2}mr^{3}x(13-x^{2}-2x^{4})+24a^{2}m^{2}r^{2}x(3-x^{2}) \\
&\qquad +2a^{4}r^{2}x(-5+4x^{2}+x^{4})+2a^{4}mrx(-3+12x^{2}+x^{4})+8a^{4}m^{2}x \\
&\qquad -a^{6}x(1+6x^{2}+x^{4}),\\[1mm]
\mathcal V_{2}&=-r^{7}(1+6x^{2}+x^{4})+2mr^{6}(1-24x^{2}-3x^{4})+8m^{2}r^{5}x^{2}(-3+2x^{2}) \\
&\qquad +a^{2}r^{5}(-1+2x^{2}-9x^{4})-2a^{2}mr^{4}(3+32x^{2}+11x^{4})+12a^{2}m^{2}r^{3}(1-14x^{2}+5x^{4}) \\
&\qquad +a^{4}r^{3}(1+6x^{2}+x^{4})+2a^{4}mr^{2}(-3+12x^{2}-23x^{4})-4a^{4}m^{2}r(1+12x^{2}+x^{4}) \\
&\qquad +a^{6}r(1-2x^{2}+9x^{4})+2a^{6}m(1+4x^{2}+x^{4}),\\[1mm]
\mathcal V_{3}&=2mr^{7}x(1+12x^{2}+x^{4})+4m^{2}r^{6}x(-1+12x^{2}-x^{4})+a^{2}r^{6}x(9-2x^{2}+x^{4}) \\
&\qquad +2a^{2}mr^{5}x(1+20x^{2}+5x^{4})+4a^{2}m^{2}r^{4}x(-9+38x^{2}-5x^{4})+a^{4}r^{4}x(1+6x^{2}+x^{4}) \\
&\qquad +2a^{4}mr^{3}x(13-8x^{2}+5x^{4})+8a^{4}m^{2}r^{2}x(2+9x^{2})+a^{6}r^{2}x(-9+2x^{2}-x^{4}) \\
&\qquad +2a^{6}mrx(-3+x^{4})-a^{8}x(1+6x^{2}+x^{4}),\\[1mm]
\mathcal V_{4}&=-24m^{2}r^{7}x^{4}-a^{2}r^{7}(1+6x^{2}+x^{4})+2a^{2}mr^{6}(1-12x^{2}-3x^{4}) \\
&\qquad +8a^{2}m^{2}r^{5}x^{2}(9-11x^{2})+2a^{4}r^{5}(1+4x^{2}-5x^{4})+4a^{4}mr^{4}(-2-13x^{2}+x^{4}) \\
&\qquad +4a^{4}m^{2}r^{3}(1-6x^{2}-5x^{4})+3a^{6}r^{3}(1+6x^{2}+x^{4})+2a^{6}mr^{2}(-1+2x^{2}-15x^{4}) \\
&\qquad +4a^{8}rx^{2}(1+3x^{2}),\\[1mm]
\mathcal V_{5}&=2a^{2}mr^{7}x(1+8x^{2}+x^{4})+4a^{2}m^{2}r^{6}x(-1-8x^{2}+5x^{4})+a^{4}r^{6}x(-3-6x^{2}+x^{4}) \\
&\qquad +4a^{4}mr^{5}x(2+7x^{2}+x^{4})-2a^{6}r^{4}x(1+6x^{2}+x^{4})+2a^{6}mr^{3}x(-1+6x^{2}+5x^{4}) \\
&\qquad +a^{8}r^{2}x(1-6x^{2}-3x^{4}).
\end{align*}
The constant part of $\omega$ has been fixed so that for $A\to0$  ($\omega_{01}=0$ and $\Delta_\phi =1$) these functions reduce to those of the Kerr black hole in a swirling universe, eqs.~\eqref{eq:kerrswirlingmetric} with $\omega_{00}=0$ before demanding regularity conditions; any other choice amounts to a shift of $\omega_{01}$
and does not change the geometry. We have checked that \eqref{eq:metricAKS} with these functions solves the vacuum Einstein equations.

 \bibliographystyle{JHEP}
 \bibliography{biblio_swirling_2026}

\end{document}